\documentclass[journal]{IEEEtran}

\ifCLASSINFOpdf
  \usepackage[pdftex]{graphicx}
\else
\fi
\usepackage[cmex10]{amsmath}
\usepackage{amssymb}
\usepackage{amsmath}
\usepackage{tabularx}
\newcolumntype{C}{>{\centering}X}
\usepackage{filecontents,lipsum}
\usepackage[noadjust]{cite}
\usepackage{graphicx}
\usepackage{subfig}

\begin{document}

\title{Sharing Data Homomorphically Encrypted with Different Encryption Keys}

\author{Reda~Bellafqira,
        Gouenou~Coatrieux,
        Dalel~Bouslimi,
        Gw\'enol\'e~Quellec
        and~Michel~Cozic
\thanks{R. Bellafqira, G. Coatrieux and  D. Bouslimi are with the Institut Mines-Telecom, Telecom Bretagne, Unité INSERM 1101 Latim, 29238 Brest Cedex, France (e-mail: reda.bellafqira@telecom-bretagne.eu; gouenou.coatrieux @telecom-bretagne.eu; dalel.bouslimi@telecom-bretagne.eu).}
\thanks{G. Quellec is with Inserm, UMR 1101, F-29200 Brest, France (e-mail: gwenole.quellec@inserm.fr).}
\thanks{M. Cozic is with MED.e.COM, Plougastel Daoulas 29470, France (e-mail: mcozic@wanadoo.fr).}
\thanks{This work was supported in part by LabCom and Region Bretagne}
}

\maketitle

\begin{abstract}
In this paper, we propose the first homomorphic
based proxy re-encryption (HPRE) solution that allows different
users to share data they outsourced homomorphically encrypted
using their respective public keys with the possibility to process
such data remotely. More clearly, this scheme makes possible to
switch the public encryption key to another one without the help
of a trusted third party. Its originality stands on a method we
propose so as to compute the difference between two encrypted
data without decrypting them and with no extra communications.
Basically, in our HPRE scheme, the two users, the delegator and
the delegate, ask the cloud server to generate an encrypted noise
based on a secret key, both users previously agreed on.
Based on our solution for comparing encrypted data, the cloud computes 
in clear the differences in-between the encrypted noise and the encrypted data of the delegator, obtaining thus blinded data. By next the cloud encrypts these differences with the public key of the delegate. As the noise is also known of the delegate, this one just has to remove it to get access to the data encrypted with his public key. This solution has been experimented in the case of the sharing of images outsourced into a semihonest cloud server.

\end{abstract}


\section{Introduction}
\label{seq1}

Nowadays, Cloud computing  allows data owners to use massive data storage and large computation capabilities at a very low costs. Despite these benefits, such a data outsourcing induces important security challenges. Indeed, data owners lose the control over the pieces of information they outsource. To protect data in terms of confidentiality and privacy from unauthorized users as well as from the cloud, one common solution consists in encrypting data. However, if using encryption achieves data confidentiality, it may limit the possible reuse or processing of outsourced data as well as the sharing of data. In this work, we are interested in the sharing of data between different users who have outsourced their data encrypted with their own public keys, i.e. using some asymmetric cryptosystem. Such a kind of problem is referred as proxy re-encryption (PRE) \cite{blaze1998divertible}, where Alice (the delegator or the emitter) wants to share with Bob (the delegate or recipient) some data she previously outsourced encrypted into the cloud (the proxy). When working with asymmetric encryption, the objective of PRE is to securely enable the proxy to re-encrypt Alice's cipher-text, encrypted with her public key, into a cipher-text that can be decrypted with Bob's private key. To do so, one simple PRE solution consists in asking Alice to provide her private key to the proxy. However, this strategy imposes the proxy to be completely trusted and does not work in the case the cloud is considered as semi-honest (i.e., it will not disclose the data but will be curious). Blaze \textit{et al}. \cite{blaze1998divertible} proposed the first PRE scheme in such a semi-honest framework. This one is based on the ElGamal cryptosystem and on a set of secret pieces of information, referred as secret re-encryption key, Alice has to send to the proxy so as to make possible the change of the public key encryption (i.e., re-encrypt data with Bob's public key). One main issue of this proposal, remarked by Ateniese \textit{et al}. \cite{ateniese2006improved}, is that Blaze \textit{et al}.'s scheme is inherently bidirectional, that is to say that the re-encryption key which allows transferring cipher-texts from Alice to Bob, enables the proxy to convert all Bob's cipher-texts under Alice's public key. This is not acceptable for Bob. The main reason of this is that the re-encryption key depends on the delegate (Bob) private key. In order to solve this problem and achieve a unidirectional PRE different approaches have been proposed. The first class of methods relies on classical asymmetric encryption cryptosystems. For instance, \cite{jakobsson1999quorum} take advantage of a quorum-based protocol which stands on distributed proxies, each of them possesses a part of the data of Alice but receive a different re-encryption key independent of Bob private key. However, with this approach, the security of Alice private key is safe as long as some proxies are honest. An alternative, proposed in \cite{dodis2003proxy}, works with only one proxy where the re-encryption key provided by Alice is split into two parts, one for the proxy and the other for Bob. Unfortunately, with \cite{dodis2003proxy}, the data of Alice, she encrypted with her public-key are turned into symmetrically encrypted data and not asymmetrically with the public key of Bob. The second class regroups methods referred as identity-based proxy re-encryption (IBPRE) and was introduced by Green and Ateniese \cite{green2007identity}. Such a method mixes PRE with identity-based cryptography (IBC). In IBC, the public encryption key of one user is derived from his identity (e.g., his email address); by combining it with PRE, the emitter and the proxy just need to know the delegates' identities instead of verifying their certificates. Basically, the unidirectional propriety is achieves due to the fact the re-encryption key depends on the identity of the delegate. However, it must be known that IB-PRE suffers of the key-escrow issue (see \cite{dodis2003proxy} for more details). Most of these schemes also rely on cryptosystems which are based on bilinear pairing \cite{han2013identity, chu2007identity, matsuo2007proxy, liang2009attribute, xu2016conditional}, an application considered as a very expensive in terms of computation complexity compared to modular multiplication or exponentiation \cite{baek2005certificateless}. To overcome this issue, Deng \textit{et al}. \cite{deng2008chosen} proposed an asymmetric cross-cryptosystem re-encryption scheme instead of pairing. 
Beyond, if the above approaches allow one user to share data with another one, they do not make possible the processing of encrypted data by the cloud or proxy. This capacity is usually achieved with the help of homomorphic cryptosystems. With these ones, one can perform operations onto encrypted data with the guarantee that the decrypted result equals the one carried out onto un-encrypted data \cite{rivest1978data}. The first homomorphic based PRE attempt  has been proposed by Bresson \textit{et al}. in \cite{bresson2003simple}, using the Paillier cryptosystem \cite{paillier1999public}. However, even though their solution makes possible data sharing, it cannot be seen as a pure proxy re-encryption scheme. Indeed, data are not re-encrypted with the public key of the delegate. If this one wants to ask the cloud to process the data he receives from Alice, he has: i) first to download Alice data, ii) decrypt them based on some secret pieces of information provided by Alice; iii) re-encrypt them with his public key and send them back to the cloud. There is thus still a need for a homomorphic based PRE.

In this work, we propose the first homomorphic proxy re-encryption scheme which does not require the delegate to re-upload the data another user has shared with him. It is based on the Paillier cryptosystem. It can be roughly summarized as follows. Bob and Alice agree on a secret key; key Alice sends Paillier encrypted to the cloud.  The cloud uses this key so as to generate a Paillier encrypted random sequence with the help of a secure linear congruential generator (SLCG) we propose and which works in the Paillier encrypted domain. All computations are conducted by the cloud server. This SLCG provides a sequence of Paillier encrypted random numbers. Based on a fast and new solution we propose so as to compute the difference in-between Paillier encrypted data, the cloud: i) computes in clear the difference between this encrypted random sequence and the encrypted data of Alice and, ii) encrypts this sequence of differences with the public key of Bob. Then, Bob just has to ask the cloud to remove the noise from the encrypted data in order to get access to the data Alice wants to share with him and process them in an outsourced manner if he wants.

The rest of the paper is organized as follow. In Section \ref{seq2}, we come back on the definition of Paillier cryptosystem and show how to use it in order to: i) quickly compute the difference between Paillier encrypted data; and ii) implement a secure  linear congruential generator so as to generate an encrypted random sequence. Section \ref{seq3} describes the overall architecture of our Homomorphic PRE solution (HPRE) in the case of the sharing of images. Performance of the proposed solution is given in Section \ref{seq4}. Conclusions are given in Section \ref{seq6}. 

\section{Processing Paillier Encrypted Data}
\label{seq2}
In this section, we first introduce the Paillier cryptosystem as well as a new way to compute the difference between Paillier encrypted data before presenting a secure linear congruential generator (LCG) implemented in the Paillier encrypted domain so as to generate an encrypted pseudo random sequence of integers.

\subsection{Paillier cryptosystem}
\label{sseq2:1}
We opted for the asymmetric Paillier cryptosystem because of its additive homomorphic property \cite{paillier1999public}. In this work, we use a fast version of it defined as follows. Let $((g,K_p), K_s)$ be the public/private key pair, such as:
	\begin{equation}
      K_p = pq \quad and \quad K_s= (p-1)(q-1)
    \label{eq1}
	\end{equation}
where $p$ and $q$ are two large prime integers. $\mathbb{Z}_{K_p}= \{0, 1,..., K_p-1\}$ and  $\mathbb{Z}_{K_p}^*$ denotes the integers that have multiplicative inverses modulo $K_p$. We select $g\in\mathbb{Z}_{K_p^2}^*$ such as:
	\begin{equation}
	\frac{g^{K_s}-1 \, mod\, K_p^2}{K_p} \in \mathbb{Z}_{K_p}^*
    \label{eq2}
	\end{equation}
The Paillier encryption of a plain-text $m\in\mathbb{Z}_{K_p}$ into the cipher-text $c\in\mathbb{Z}_{K_p^2}^*$ using the public key $K_p$ is given by
\begin{equation}
 		c= E[m,r]= g^m r^{K_p} \mod K_p^2
        \label{eq3}
\end{equation}
where $r \in \mathbb{Z}_{K_p}^*$  is a random integer associated to $m$ making the Paillier cryptosystem probabilistic or semantically secure. More clearly, depending on the value of $r$, the encryption of the same plain-text message will yield to different cipher-texts even though the public encryption key is the same. Notice that it is possible to get a fast version of~\eqref{eq3} by fixing $g=1+K_p$ without reducing the algorithm security. By doing so, the encryption of $m$ into $c$ requires only one modular exponentiation and two modular multiplications
\begin{equation}
c=E[m,r]=(1+mK_p)r^{K_p} \mod K_p^2
\label{eq4}
\end{equation}
As we will see in section \ref{sseq2:2}, this property will be of importance for the computation of the difference between Paillier encrypted data.\\
Based on the assumption $g=1+K_p$, the decryption of $c$ using the private Key $K_s$ is such as
\begin{equation}
m=\frac{(c^{K_s}-1)K_s^{-1} \mod K_p^2}{K_p} \mod K_p
\label{eq5}
\end{equation}
If we consider two plain-texts $m_1$ and $m_2$, the additive homomorphic property of the Paillier cryptosystem allows linear operations on encrypted data like addition and multiplication, ensuring that
\begin{equation}
E[m_1,r_1]E[m_2,r_2] = E[m_1+m_2, r_1r_2]
\label{eq6}
\end{equation}

\begin{equation}
E[m_1,r_1]^{m_2} = E[m_1 m_2, r_1^{m_2}]
\label{eq7}
\end{equation}

\subsection{Computing the difference in-between encrypted data}
\label{sseq2:2}
In this work, we propose a solution that allows the calculation by one server of the difference between two Paillier encrypted data. More clearly if $a$ and $b$ are two integers, we want to compute their difference $a-b$ from their encrypted versions. 
Let us consider a user-server relationship where the server has two cipher-texts $E_{K_p}[a,r]$ and $E_{K_p}[b,r]$ encrypted by the user. It is important to notice that to make such computation possible; the two cipher-texts have to be encrypted with the same random value $r$. Under this constraint, one can directly derive the difference $d$ between $a$ and $b$ from $E_{K_p}[a,r]$ and $E_{K_p}[b,r]$ by taking advantage of the fast Paillier cryptosystem assumption, i.e. $g=1+K_p$, as follows

		$$\begin{array}{ccc}
		 
		d & = & D(a,b) =D^e(E_{K_p}[a, r], E_{K_p}[b, r]) \\[.3cm] 
		 
		 & = & \frac{E_{K_p}[a,r]E_{K_p}[b,r]^{-1}-1 \mod K_p^2}{K_p} \mod K_p \\[.3cm]  
		 
		 & = & \frac{g^a r g^{-b}r^{-1}-1 \mod K_p^2}{K_p} \mod K_p \\[.3cm]  
		
		 & = & \frac{g^{a-b} -1 \mod K_p^2}{K_p} \mod K_p \\[.3cm]  
		 
		d & = & a- b \mod K_p 
		\end{array}$$
\begin{equation}
\label{eq8}
\end{equation}
where $D$ and $D^e$ denote the two functions that allows computing the difference $d$ in the clear and Paillier encrypted domain, respectively. Notice that knowing the difference $d$ between $a$ and $b$ gives no clues about the values of $a$ and $b$, respectively.

\subsection{Secure Linear Congruential Generator}
\label{sseq2:3}
 As stated in the introduction, our HPRE scheme will require the cloud to securely generate a pseudo random sequence that is to say a Paillier encrypted random sequence of integers.  
The generator we propose to secure is LCG \cite{l1999tables} (Linear Congruential Generator). This one is based on congruence and a linear functions; functions that can be easily implemented in the Paillier encrypted domain. 
In the clear domain, LCG works as follows 
\begin{equation}
	X_{n+1} = a X_n + c \mod m 
\label{eq9}
\end{equation}
where: $X_n$ is the $n^{th}$ random integer value of the LCG sequence; $a$ is a multiplier; $c$ is an increment; $m$ is the modulo; and, $X_0$ the initial term, also called the seed or the secret LCG key, one needs to know so as to re-generate a random sequence. The security of the LCG is based on the seed $X_0$. The knowledge of the parameters $a$, $c$ and $m$ does not endanger its security \cite{l1999tables}.
This random generator can be implemented into the Paillier encrypted domain, i.e. turned into a Secure LCG (SLCG), so as to generate an encrypted random sequence of integers (i.e. $\{E[X_n,r_n ]\}_{n=0...N-1}$) in the following way :
\begin{equation}
E[X_{n+1},r_{n+1}]=E[X_n,r_n ]^a E[c,r_c] =E[a X_n+c,r_n^ar_c]
\label{eq10}
\end{equation}
 under the constraint however that $m$ equals the user Paillier public key $K_p$, (i.e.,  $m = K_p$, see ~\eqref{eq1}). 
If the increment as well as all terms of the sequence are encrypted (including the LCG seed) that is not the case of the multiplier $a$. However, this does not reduce the security of our system as the parameter $a$ is not supposed to be secret \cite{l1999tables}. 
It is important to notice that, in our SLCG, a recursive relation exists between the random integers $r_n$ which ensure the semantic security of the Paillier cryptosystem. Derived from ~\eqref{eq10}, this one is such as:
\begin{equation}
			r_{n+1}= r_n^a r_c   
\label{eq11}
\end{equation}  
 where $r_c$ is the random variable used to encrypt the increment. $r_0$ is the random value associated to the seed $X_0$. This recursive relationship will be considered in Section \ref{sseq3:2} so as to allow data exchange between two different users.
 
\begin{figure}[!t]
\centering
\includegraphics[scale=.5]{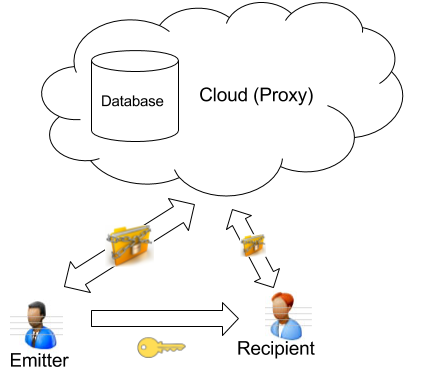} 
\caption{General framework for data sharing through public-cloud}
\label{fig1}
\end{figure}

\begin{figure*}[t]
    \begin{center}
        \includegraphics[width=0.9\textwidth]{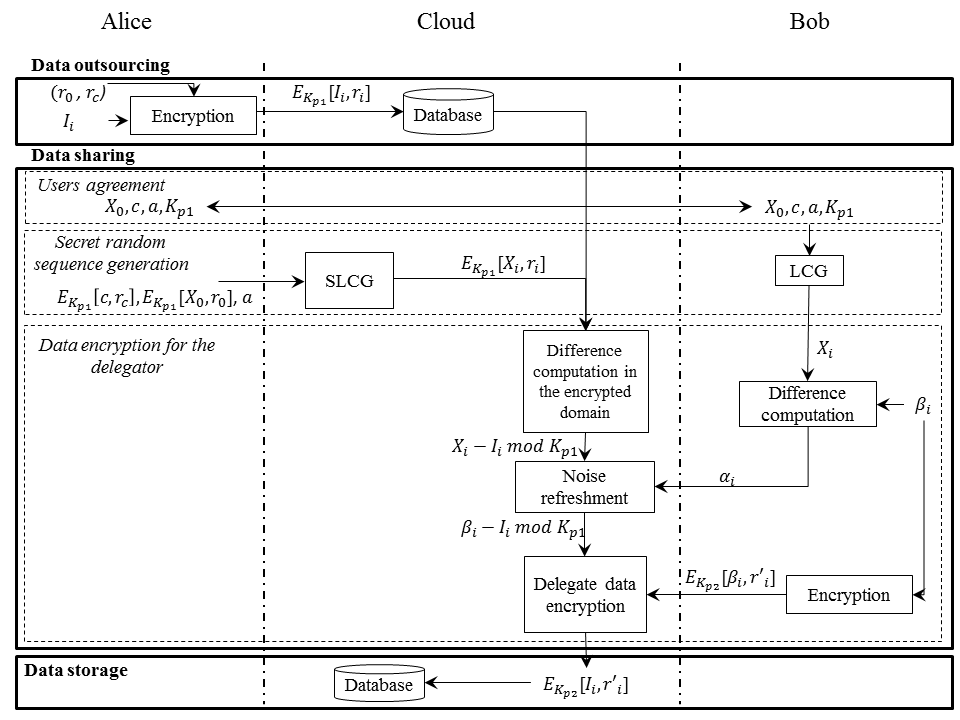}
    \end{center}
\caption{Main steps of our HPRE for an image sharing}
\label{fig2} 
\end{figure*} 
\section{Sharing outsourced encrypted data} 
\label{seq3}
 In this Section, we first refine the data exchange framework we consider and its basic security assumptions. We then present our Homomorphic based Proxy Re-Encryption scheme (HPRE). 
 \subsection{Data exchange scenario in outsourced environment}
\label{sseq3:1} 
  Fig. \ref{fig1} illustrates the general data exchange framework we consider where a data owner (the emitter or the delegator) has \textit{a priori} stored his data into a public cloud in an asymmetrically encrypted form; data he wants to share with another user (the recipient or the delegate). We further assume a semi-honest cloud server. This one honestly stores encrypted data uploaded by the users and responds to their requests. If the server does not disclose data to any parties who fail to prove ownership or access rights, it is however curious and may try to infer information about the content of users' data or about their private keys. One last assumption is that all communications between the server and the users are protected with the help of the Paillier cryptosystem. Eavesdroppers cannot infer messages being transmitted. 
As stated previously, our objective is to allow them to share some data under the constraint the delegator does not have to download his data, re-encrypt them with the public key of the delegate and upload them into the cloud. We also want this process conducted by the cloud (proxy) without giving the delegator private key as well as with very few communications in-between the delegator, the proxy and the delegate. In our idea, if one user wants to share data with several users at once, all of them will have to agree on a single secret with the delegator. 

\subsection{Secure data exchange between users}
\label{sseq3:2} 
Let us thus consider that Alice (the delegator) wants to share with Bob (the delegate) a set of data she is the owner of. These data could be a set of integer values like for instance a gray-scale image $I$, the $N$ pixels of which $I=\{I_i\}_{i=0..N-1}$   are encoded on $b$ bits. 
As stated previously, it is assumed that Alice has already outsourced an image into the cloud by Paillier encrypting its pixels independently with her public key $K_{p1}$, such as (see in Fig. \ref{fig2} – Data outsourcing step)
\begin{equation}
I_i^e=E_{K_{p1}}[I_i,r_i]
\label{eq12}
\end{equation}
where $r_i$ is the random value associated to the $i^{th}$ pixel $I_i$ of $I$, $I_i^e$ is the encrypted version of $I_i$. As we will see in the sequel, our HPRE procedure imposes a constraint on the way Alice generates the random values $\{r_i\}_{i=0..N-1}$. These ones should satisfy ~\eqref{eq11}, that is to say that for one file Alice stores into the cloud, she has to memorize the first random value $r_c$ and $r_0$ she used when she encrypted the first pixel of her image (or of any files she stored), $I_0^e=E_{K_{p1}}[I_0,r_0]$.

In order to share this encrypted image with Bob, the public Paillier encryption key of whom is $K_{p2}$, we propose the following HPRE procedure also depicted in Fig. \ref{fig2} 

\begin{enumerate}
\item \textbf{User agreement for data exchange} - In this first tsep, Bob and Alice have to agree on the exchange by defining the LCG parameters, in other words: the secret key $X_0$, the multiplier $a$ and the increment $c$. Let us recall that knowing $c$ and $a$ is not critical from a security point of view (see Section \ref{sseq2:3}).
\item \textbf{Secret random sequence generation} - Alice encrypts $X_0$ and $c$ under her public key $K_{p1}$: $E_{K_{p1}}[X_0,r_0]$  $E_{K_{p1}}[c,r_c]$, and sends them to the cloud. Notice that $X_0$ is encrypted with the same random integer $r_0$ Alice used to encrypt the first pixel of her image (see above). She also sends the multiplier $a$. Based on these pieces of information, the cloud generates the secret random sequence $X^e=\{X_i^e=E_{K_{p1}}[X_i,r_i]\}_{i=0..N-1}$  using ~\eqref{eq10}. 
\item \textbf{Data encryption for the delegator} - This procedure relies on different stages: i) the computation of differences between the encrypted data of Alice $(I^e)$ and the secret random sequence $(X^e)$; ii) the encryption of this differences with the public key of Bob $K_{p2}$.
\begin{enumerate}
\item 	\textit{Difference computation} - since $X_i^e$ and $I_i^e$ have been encrypted with the same public key $K_{p1}$ and the same random values $r_i$ (see above), the cloud computes their difference $D_i$ as exposed in see Section \ref{sseq2:2}, that is to say 
\begin{equation}
\begin{small}
   \begin{array}{ccc}
D_i & = &D(X_i,I_i )=D^e(E_{K_{p1}}[X_i,r_i],E_{K_{p1}}[I_i,r_i ]) \\ 
• & = & X_i-I_i  \mod  K_{p1} 
\end{array}
   \end{small}   
\label{eq13}
\end{equation}            
Even though the cloud knows $D=\{D_i\}_{i=0...N-1}$, it cannot deduce the value of $I_i$ and $X_i$.
\item 	\textit{Data encryption for the delegator} - From this stand point, one may think the cloud just has to encrypt $D$ with the public key of Bob, $K_{p2}$, and then remove the noise so as so to give him access to the data. This is possible under the constraint $D_i  \mod K_{p1} = D_i  \mod K_{p2}$ which is achieved when $0<D_i<min(K_{p1}, K_{p2})$. Unfortunately, this constraint is hard to satisfy because of the SLCG the output amplitude of which can not be controlled simply. To overcome this issue, our HPRE includes a "noise refreshment procedure" (see Fig. \ref{fig2}) before encrypting the data with the public key of Bob. 

\begin{itemize}
\item Noise refreshment
\end{itemize}
 To refresh the noise, Bob first generates on his side the sequence $\{X_i\}_{i=0..N-1}$, using an LCG parameterized as the SLGC of the cloud. 
He also produces a second noise $\{\beta_i\}_{i=0..N-1}$ such as: 
\begin{equation}
		2^b-1<\beta_i<\min(K_{p1},K_{p2})		
\label{eq14}
\end{equation}			
where $b$ is the number of bits on which is encoded the pixel values of the image of Alice. Under such a constraint: we ensure:  $\beta_i \mod K_{p1}= \beta_i  \mod K_{p2}$ and $\beta_i-I_i  \mod K_{p1}= \beta_i-I_i  \mod K_{p2}$. 
Then Bob sends to the cloud $\{E_{K_{p2}}[\beta_i,r_i']\}_{i=0...N-1}$ and $\{\alpha_i=\beta_i-X_i  \mod K_{p1}\}_{i=0..N-1}$. Where $r_i'$ is a random value defined by Bob. 

On its side, in order to remove the noise $\{X_i\}_{i=0...N-1 }$, the cloud computes
\begin{equation}
G_i=\alpha_i+D_i \mod K_{p1}=\beta_i-I_i  \mod K_{p1}
\label{eq15}
\end{equation}
Then it encrypts $\{G_i\}_{i=0..N-1}$ with the public key of Bob
\begin{equation}
\{E_{K_{p2}}[G_i,r_i'']=E_{K_{p2}}[\beta_i-I_i,r_i'']\}_{i=0..N-1} 	
\label{eq16}
\end{equation} 
Finally, in order to remove the noise $\beta_i$ from of the data of Bob, the server computes
\begin{equation}
E_{K_{p2}}[I_i,r_i' r_{i}^{''-1}]=E_{K_{p2}}[\beta_i,r_i']E_{K_{p2}}[\beta_i-I_i,r_i'']^{-1}
\label{eq17}
\end{equation}
\end{enumerate}
At the end of this procedure, Bob has on the cloud the image of Alice encrypted with his own public key.
\end{enumerate}	
As depicted, this system allows the data exchange between Alice and Bob, without extra-communication between the cloud and Alice, and the downloading of data by Bob. It is also possible to notice that the access to the shared data is based on the knowledge  of the secret SLCG key $X_0$ generated by Alice in agreement with Bob. Because our scheme is based on homormophic encryption, data can be by next processed by the cloud without endangering data confidentiality. 

\begin{table*}[t]
 	\renewcommand{\arraystretch}{1.5}  
    \begin{center}
        \begin{tabular}{|c|c|c|c|}
			\hline
				Entities & Delegator (Alice) & Proxy (Cloud) & Delegate (Bob) \\
			\hline
				Time computation (sec) & 0.002 & 90 & 30\\
			\hline
				Encrypted data volume of & 0 & 22986753 & 2048 \\
			\hline
		\end{tabular}
    \end{center}
\caption{Amount of information stored (in bits) as well as the corresponding computation time that each entity needs (Alice, Bob and the cloud) for sharing an image of $92\times 122$ pixels}
\label{tab1} 
\end{table*}

\begin{figure}[!t]
\centering
\includegraphics[scale=.5]{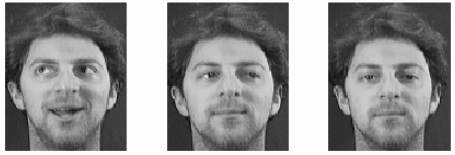} 
\caption{Samples of face database}
\label{fig3}
\end{figure}

\section{Experimental results} 
\label{seq4} 
The previous solution was experimented in the case of the sharing of uncompressed images between two users. These images are issued from the Olivetti Research Laboratory of Cambridge, UK. It contains $400$ images of $8$ bit encoded of $92\times 122$ pixels. Some samples of our image test set are given Fig. \ref{fig3}. These images were encrypted with Paillier public keys of more than $1024$ bits in order to provide a high level of security. 
Performance of our scheme are evaluated in terms of storage and computation complexity. Our HPRE was implemented in C/C++ with GMP library and all experiments were conducted using a machine equipped with $23$GB RAM running on Ubuntu $14.04$ LTS.
	\begin{itemize}
	\item Storage complexity:
	\end{itemize}	 
Assuming that images are Paillier encrypted with a key of $1024$ bits, one encrypted image needs $2,7Mo$ so as to be stored into the cloud. 
For one image the delegator (Alice) outsources, she only has to store on her side the random values $r_0$ and $r_c$. However this is not obligation. Indeed, based on the fact she knows both her public and private keys, she just has to download the encrypted seed $E_{K_{p1}}[X_0,r_0]$  and the encrypted increment $E_{K_{p1}}[c,r_c]$, to get access to these random values (i.e. $r_c$ and $r_0$). 
During an image exchange, the delegator sends the encrypted seed $E_{K_{p1}} [X_0,r_0]$, the encrypted increment $E_{K_{p1}}[c,r_c]$ and the multiplier $a$. This amount of data is bounded by $O(log_2(K_{p1}^2))$. For a key of $1024$ bits, it is closed to $2048$ bits. On its side, the delegate (Bob) has to store $X_0$, the secret key of the LCG, but only for one session of data exchange. 
	\begin{itemize}
	\item Computation complexity:
	\end{itemize}
On the delegator side, the computation complexity is limited to the encryption of the SLCG parameters (i.e. $X_0$, $c$). Such a complexity is independent of the image's size.
Regarding the cloud, this one has to compute: the secret random sequence, compute the difference between the encrypted date of Alice with this random sequence, refresh the noise based on the inputs of Bob, encrypt the result with the public key of Bob and finally remove the noise. For an image of $N$ pixels, the secret random sequence generation is equivalent to $N$ encryptions. It is the same for the computation of the differences $\{D_i\}_{i=0...N-1}$. As described above, the noise refreshment procedure consists in modular additions. We consider its complexity negligible compared to encryption operations. The last step, the encryption of the differences $\{G_i\}_{i=0...N-1}$ is made of $N$ encryptions. As a consequence, the computation complexity for the cloud is bounded by $O(3\times N)$ encryptions.

The delegate computation complexity is attached to the noise refreshment procedure. He has to generate a LCG noise (i.e. $\{X_i\}_{i=0...N-1}$), a task the complexity of which is negligible compared to the $N$ encryptions of the second noise (i.e. $\{\beta_i\}_{i=0...N-1}$) he also produces and that he next sends to the cloud. The computation complexity of the delegate is thus of $N$ encryptions. 
We provide in Table \ref{tab1} the amount of data that each entities has to store as well as the computation time required in the case of sharing images of our data set. Our HPRE scheme takes about $1'30$ minutes so as to share an image with a standard computer.

\section{Conclusion}
\label{seq6}
In this paper, we proposed the first homomorphic proxy re-encryption scheme. Its originality stands on a solution we propose so as to compute the difference of data encrypted with the fast version of the Paillier cryptosystem. It takes also advantage of a secure linear congruential generator we implemented in the Paillier encrypted domain. This one drastically reduces the computation complexity of the cloud and delegator. Furthermore, this solution doesn't need extra communication between the cloud and the delegator, i.e. the data owner. Moreover, since the data are homomorphically encrypted, it is possible to process outsourced data while ensuring their confidentiality. Our HPRE was implemented in the case of the sharing of uncompressed images stored in the cloud showing good time computation performance. Our scheme is not limited to images and can be used with any kinds of data.


\bibliographystyle{IEEEtran}
\bibliography{biblio}

\begin{thebibliography}{10}
\providecommand{\url}[1]{#1}
\csname url@samestyle\endcsname
\providecommand{\newblock}{\relax}
\providecommand{\bibinfo}[2]{#2}
\providecommand{\BIBentrySTDinterwordspacing}{\spaceskip=0pt\relax}
\providecommand{\BIBentryALTinterwordstretchfactor}{4}
\providecommand{\BIBentryALTinterwordspacing}{\spaceskip=\fontdimen2\font plus
\BIBentryALTinterwordstretchfactor\fontdimen3\font minus
  \fontdimen4\font\relax}
\providecommand{\BIBforeignlanguage}[2]{{%
\expandafter\ifx\csname l@#1\endcsname\relax
\typeout{** WARNING: IEEEtran.bst: No hyphenation pattern has been}%
\typeout{** loaded for the language `#1'. Using the pattern for}%
\typeout{** the default language instead.}%
\else
\language=\csname l@#1\endcsname
\fi
#2}}
\providecommand{\BIBdecl}{\relax}
\BIBdecl

\bibitem{blaze1998divertible}
M.~Blaze, G.~Bleumer, and M.~Strauss, ``Divertible protocols and atomic proxy
  cryptography,'' in \emph{International Conference on the Theory and
  Applications of Cryptographic Techniques}.\hskip 1em plus 0.5em minus
  0.4em\relax Springer, 1998, pp. 127--144.

\bibitem{ateniese2006improved}
G.~Ateniese, K.~Fu, M.~Green, and S.~Hohenberger, ``Improved proxy
  re-encryption schemes with applications to secure distributed storage,''
  \emph{ACM Transactions on Information and System Security (TISSEC)}, vol.~9,
  no.~1, pp. 1--30, 2006.

\bibitem{jakobsson1999quorum}
M.~Jakobsson, ``On quorum controlled asymmetric proxy re-encryption,'' in
  \emph{International Workshop on Public Key Cryptography}.\hskip 1em plus
  0.5em minus 0.4em\relax Springer, 1999, pp. 112--121.

\bibitem{dodis2003proxy}
Y.~Dodis and A.~Ivan, ``Proxy cryptography revisited,'' in \emph{Proceedings of
  the Tenth Network and Distributed System Security Symposium}, 2003, pp.
  514--532.

\bibitem{green2007identity}
M.~Green and G.~Ateniese, ``Identity-based proxy re-encryption,'' in
  \emph{Applied cryptography and network security}.\hskip 1em plus 0.5em minus
  0.4em\relax Springer, 2007, pp. 288--306.

\bibitem{han2013identity}
J.~Han, W.~Susilo, and Y.~Mu, ``Identity-based data storage in cloud
  computing,'' \emph{Future Generation Computer Systems}, vol.~29, no.~3, pp.
  673--681, 2013.

\bibitem{chu2007identity}
C.-K. Chu and W.-G. Tzeng, ``Identity-based proxy re-encryption without random
  oracles,'' in \emph{International Conference on Information Security}.\hskip
  1em plus 0.5em minus 0.4em\relax Springer, 2007, pp. 189--202.

\bibitem{matsuo2007proxy}
T.~Matsuo, ``Proxy re-encryption systems for identity-based encryption,'' in
  \emph{International Conference on Pairing-Based Cryptography}.\hskip 1em plus
  0.5em minus 0.4em\relax Springer, 2007, pp. 247--267.

\bibitem{liang2009attribute}
X.~Liang, Z.~Cao, H.~Lin, and J.~Shao, ``Attribute based proxy re-encryption
  with delegating capabilities,'' in \emph{Proceedings of the 4th International
  Symposium on Information, Computer, and Communications Security}.\hskip 1em
  plus 0.5em minus 0.4em\relax ACM, 2009, pp. 276--286.

\bibitem{xu2016conditional}
P.~Xu, T.~Jiao, Q.~Wu, W.~Wang, and H.~Jin, ``Conditional identity-based
  broadcast proxy re-encryption and its application to cloud email,''
  \emph{IEEE Transactions on Computers}, vol.~65, no.~1, pp. 66--79, 2016.

\bibitem{baek2005certificateless}
J.~Baek, R.~Safavi-Naini, and W.~Susilo, ``Certificateless public key
  encryption without pairing,'' in \emph{International Conference on
  Information Security}.\hskip 1em plus 0.5em minus 0.4em\relax Springer, 2005,
  pp. 134--148.

\bibitem{deng2008chosen}
R.~H. Deng, J.~Weng, S.~Liu, and K.~Chen, ``Chosen-ciphertext secure proxy
  re-encryption without pairings,'' in \emph{International Conference on
  Cryptology and Network Security}.\hskip 1em plus 0.5em minus 0.4em\relax
  Springer, 2008, pp. 1--17.

\bibitem{rivest1978data}
R.~L. Rivest, L.~Adleman, and M.~L. Dertouzos, ``On data banks and privacy
  homomorphisms,'' \emph{Foundations of secure computation}, vol.~4, no.~11,
  pp. 169--180, 1978.

\bibitem{bresson2003simple}
E.~Bresson, D.~Catalano, and D.~Pointcheval, ``A simple public-key cryptosystem
  with a double trapdoor decryption mechanism and its applications,'' in
  \emph{International Conference on the Theory and Application of Cryptology
  and Information Security}.\hskip 1em plus 0.5em minus 0.4em\relax Springer,
  2003, pp. 37--54.

\bibitem{paillier1999public}
P.~Paillier, ``Public-key cryptosystems based on composite degree residuosity
  classes,'' in \emph{International Conference on the Theory and Applications
  of Cryptographic Techniques}.\hskip 1em plus 0.5em minus 0.4em\relax
  Springer, 1999, pp. 223--238.

\bibitem{l1999tables}
P.~L’ecuyer, ``Tables of linear congruential generators of different sizes
  and good lattice structure,'' \emph{Mathematics of Computation of the
  American Mathematical Society}, vol.~68, no. 225, pp. 249--260, 1999.

\end{thebibliography}

\end{document}